\documentclass[paper]{ieice}
\usepackage[dvips]{graphicx}
\usepackage[fleqn]{amsmath}
\usepackage{subfigure}
\usepackage[varg]{txfonts}
\usepackage{cite}
\usepackage{epsfig,graphics,mathrsfs,amsmath,bm,color,yfonts,txfonts}
\usepackage{subfigure}
\usepackage[imp,figs]{optional}

\field{EB}
\vol{}
\no{}

\title{An Efficient Algorithm for Weighted Sum-Rate Maximization in Multicell OFDMA Downlink}
\authorlist{
\authorentry[contact-h25e@hanase.kuee.kyoto-u.ac.jp]{Mirza Golam Kibria}{s}{kyoto-u}
\authorentry{Hidekazu MURATA}{e}{kyoto-u}
\authorentry{Susumu YOSHIDA}{f}{kyoto-u}
}
\affiliate[kyoto-u]{The authors are with the Graduate School of Informatics, Kyoto University, Kyoto-shi, 606-8501 Japan.}

\received{2013}{3}{15}
\revised{2013}{7}{27}

\begin{document}
\maketitle

\begin{summary}
This paper considers coordinated linear precoding for rate optimization in downlink multicell, multiuser orthogonal frequency-division multiple access networks. We focus on two different design criteria. In the first, the weighted sum-rate is maximized under transmit power constraints per base station. In the second, we minimize the total transmit power satisfying the signal-to-interference-plus-noise-ratio constraints of the subcarriers per cell. Both problems are solved using standard conic optimization packages. A less complex, fast, and provably convergent algorithm that maximizes the weighted sum-rate with per-cell transmit power constraints is formulated. We approximate the non-convex weighted sum-rate maximization (WSRM) problem with a solvable convex form by means of a sequential parametric convex approximation approach. The second-order cone formulations of an objective function and the constraints of the optimization problem are derived through a proper change of variables, first-order linear approximation, and hyperbolic constraints transformation. This algorithm converges to the suboptimal solution while taking fewer iterations in comparison to other known iterative WSRM algorithms. Numerical results are presented to demonstrate the effectiveness and superiority of the proposed algorithm.
\end{summary}
\section{Introduction}
The multiantenna Gaussian broadcast channel (BC) has recently been a subject of considerable interest and research, primarily due to its inherent feature of realizing multi-input multi-output (MIMO) spatial multiplexing benefits, even when the mobile devices have a single antenna each \cite{Caire}. Costa precoding or dirty paper coding (DPC) \cite{Costa}, a nonlinear interference cancellation technique, is known to be a capacity-achieving strategy for BC. However, DPC implementation demands substantial extra complexity at both the base station (BS) and receivers. To the contrary, linear precoding, a suboptimal precoding technique, can achieve a large fraction of DPC capacity at significantly lower complexity without compromising the multiplexing gain. Moreover, a linear precoder can be designed so as to deal with the weighted sum-rate maximization (WSRM) problem under per-BS power constraints \cite{Venturino}.

The WSRM problem with BS transmit power constraints is non-convex and NP-hard\cite{Venturino,Wang}, even for single antenna users. As a result, finding the global optimal solution for this non-convex problem is very difficult. Though the beamforming designs presented in \cite{Joshi, Liu} achieve optimal capacity, these optimal techniques may be practically inapplicable because the computational complexity increases exponentially with the optimization problem size. Therefore, suboptimal design with less complex signal processing is very important. Beamforming designs achieving necessary optimality conditions have been thoroughly studied in \cite{Venturino,Wang}. Surprisingly, it has been proved numerically in \cite{Joshi} that the performances of suboptimal methods achieving the necessary optimality conditions are indeed very close to the optimal beamforming performance.

An iterative coordinated beamforming design based on Karush-Kuhn-Tucker (KKT) optimality conditions has been proposed in \cite{Venturino}, which is not provably convergent.
A solution for the WSRM optimization problem with an alternating maximization (AM) algorithm is presented in \cite{Wang}, which relies on alternate updating between the beamforming vectors and a closed-form posterior conditional probability. In \cite{Christensen,Sun,Shi}, the authors solved the WSRM problem with alternating optimization, establishing a relationship between weighted sum-rate and weighted minimum mean-square error (WMMSE). Iterative and discrete power-control-based solutions for WSRM have been discussed in \cite{Zhang1}. However, these iterative WSRM optimization designs exhibit relatively slower convergence rates.

In this paper, we formulate and propose a less complex and faster convergent solution for the WSRM problem for multicell, multiuser orthogonal frequency-division multiple access (OFDMA) systems. Our beamforming design is based on a sequential parametric convex approximation (SPCA) approach explored in \cite{Beck}. This SPCA-based WSRM (SPCA-WSRM) optimization converges to the local optimal solution within a few iterations by iteratively approximating the non-convex WSRM problem with a solvable convex form. In particular, the WSRM problem is approximated as a second-order cone program (SOCP) \cite{Lobo} with proper change of variables, introducing additional optimization variables, applying first-order linear approximation and hyperbolic constraints transformations.

Power optimization under signal-to-interference-plus-noise-ratio (SINR) constraints has been solved optimally by various approaches. The downlink beamforming problem under per-subcarrier SINR constraints has been solved via equivalent uplink problem exploiting the uplink-downlink duality in \cite{Rashid,Dahrouj}. In \cite{Nguyen}, a game theoretical approach is used. In \cite{Rashid2}, the authors considered power allocation and beamforming as a joint problem and efficiently solved the joint problem under SINR constraints, iteratively updating the uplink beamformer and transmit power. We refer to this method as JBPA. The optimality of the uplink-downlink duality-based approach in a single cell multiuser case has been proved in \cite{Visotsky,Schubert2}. The authors of \cite{Bengtsson} formulated the downlink beamforming problem as a convex semi-definite program (SDP) using a constraint relaxation technique. More recently, in \cite{Wiesel}, SOCP formulation was used to find the global optimal solution for this power optimization problem. Inter-cell interference cancellation through user scheduling in downlink beamforming has been performed in \cite{Ren}. Most of these works are based on single-carrier systems. 

In this paper, we formulate the power optimization problem under SINR constraints in multicell multicarrier downlink beamforming using SOCP. Because the structure of the optimization problem is convex \cite{Rashid, Dahrouj}, no approximation is required and the global optimal solution is guaranteed. The convergence behaviors of SOCP-based optimization are compared with other well-known iterative methods.

The reminder of the paper is organized as follows. The multicell, multiuser OFDMA network model and WSRM optimization framework are presented in Section 2. Section 3 explains the process of sequential convex approximation of the non-convex optimization problem. We illustrate the power optimization problem under SINR constraints in Section 4. In Section 5, we discuss the simulation parameters and numerical results found in this work. Section 6 concludes the paper.\\
\textit{Notations:} Superscripts $(\cdot)^{\rm{H}}$ and $(\cdot)^{\rm{T}}$ denote the Hermitian transpose and transpose operations, respectively. Gaussian distribution of real/complex random variables with mean $\mu$ and variance $\sigma^2$ is defined as $\mathcal{RN} (\mu,\sigma^2)$/$\mathcal{CN} (\mu,\sigma^2)$. Boldface lower-case/upper-case letters defines a vector/matrix. Operator $\mathrm{vec}(\cdot)$ stacks all the elements of the argument into a column vector, and $\mathrm{diag}(\cdot)$ puts the diagonal elements of a matrix into a column vector. $\mathbb{R}$/$\mathbb{C}$ defines a real/complex space. $\Re\{\cdot\}$ and $\Im\{\cdot\}$ denote the real and imaginary part of the arguments, respectively. $|\cdot|$ and $||\cdot||_2$ refer to the absolute value and $l_2$ norm of the arguments, respectively.
\section{Problem Formulation}
\subsection{System Model}
We consider an interference-limited cellular system of $N_{\mathrm{cells}}$ cells with $N_{\mathrm{users/cell}}$ users per cell. An OFDMA multiplexing scheme with $N_{\mathrm{sub}}$ subcarriers over a fixed bandwidth is employed, where the subcarrier assignments are non-overlapping\footnote{Note that the non-overlapping subcarrier allocation in each cell does not limit the applicability of our proposed algorithm to the case of multiple active users in one subcarrier in one cell. Because the BS has multiple transmit antennas, it can serve multiple users in each subcarrier at the same time. Then the user experiences co-channel interference within its own cell.} among the users within a cell. As a result, only inter-cell interference is experienced by the users; there is no intra-cell interference. Each coordinated BS is equipped with $N_\mathrm{Tx}$ antennas, whereas the non-cooperative users have a single antenna each. The BSs are interconnected via high-capacity backhaul links, and global channel state information (CSI) is shared among them. Coordinated linear multiuser downlink precoding is employed at each BS. The downlink user scheduling is determined by the assignment function $\phi(m,n)$. The assignment of user $k$ from the $m$th BS on the $n$th subcarrier is defined as $k=\phi(m, n)$. Let the set of all the cells be denoted by $\mathcal{M}=\{1,2,\cdots,N_{\mathrm{cells}}\}$, and let $\mathcal{N}=\{1,2,\cdots,N_{\mathrm{sub}}\}$ be the set of all subcarriers. The baseband signal model for the received data of user $k$ from cell $m$ on the $n$th subcarrier is given by
\vspace{-1mm}
\begin{equation}
{{y}_{kmn}}={{\bm{h}}_{kmn}}{{\bm{f}}_{kmn}}{{d}_{kmn}}+\sum\limits_{\begin{smallmatrix}
 {m}'\in \mathcal{M}\backslash m \\
{k'}=\phi(m', n)
\end{smallmatrix}}{{{\bm{h}}_{km'n}}{{\bm{f}}_{k'{m}'n}}{{d}_{{k}'{m}'n}}}+{{z}_{kmn}},
\end{equation}
where $ \mathcal{M}\backslash m$ defines the subtraction of element $m$ from set $\mathcal{M}$. $\text{ }y_{kmn}\in \mathbb{C}$ denotes the received symbol for user $k$ with $k=\phi(m,n)$. $\bm{h}_{kmn}\in \mathbb{C}^{{1\times N_\mathrm{Tx}}} $ is the complex channel vector between user $k$ and BS $m$. The beamformer used by BS $m$ to transmit data on subcarrier $n$ is given by $\bm{f}_{kmn}\in \mathbb{C}^{{N_\mathrm{Tx}}\times 1} $. $d_{kmn}\sim{\mathcal{CN}(0,1)}$ denotes the transmitted symbol from BS $m$ to user $k$ on subcarrier $n$, and $z_{kmn}\sim{\mathcal{CN}(0,1)}$ is the additive white Gaussian noise (AWGN) at user $k$. 
\subsection{Transmit Precoding Problem}
In this paper, we emphasize the linear beamformer design for WSRM optimization in a multicell, multiuser OFDMA network. The design objective is the maximization of the weighted sum-rate under per-BS transmit power constraints. The SINR of the $k$th user from cell $m$ scheduled on subcarrier $n$ is given by
\begin{equation}
\label{gamma}
{{\gamma }_{kmn}}=\frac{{{\bm{h}}_{kmn}}{{\bm{f}}_{kmn}}\bm{f}_{kmn}^{\mathrm{H}}\bm{h}_{kmn}^{\mathrm{H}}}{1+\sum\limits_{\begin{smallmatrix}
 {m}'\in \mathcal{M}\backslash m \\
 {k'}=\phi(m', n)
 \end{smallmatrix}}{{{\bm{h}}_{km'n}}{{\bm{f}}_{k'{m}'n}}\bm{f}_{k'{m}'n}^{\mathrm{H}}\bm{h}_{k{m}'n}^{\mathrm{H}}}}.
\end{equation}
The instantaneous downlink data-rate achieved by the $k$th user from cell $m$ on subcarrier $n$ is ${{r}_{kmn}}={{\log }_{2}}(1+{{\gamma }_{kmn}})$, and the instantaneous rate over all the subcarriers is given by ${{R}_{km}}=\sum\nolimits_{n\in {{\mathcal{S}}_{km}}}{{{r}_{kmn}}}$. The summation is over all the subcarriers scheduled to user $k$ from cell $m$, i.e., $n\in {\mathcal{S}_{km}}$, where ${{\mathcal{S}}_{km}}=\left\{ n\hspace{1mm}|\hspace{1mm}k=\phi(m,n) \right\}$. Let the weight of user $k$ in cell $m$ is defined as $w_{km}$, which may correspond to the quality of the service it requests or some sort of priority in the system. Consequently, the WSRM problem is defined as
\begin{equation}
\label{optzm}
\begin{aligned}
& \underset{\bm{\mathcal{F}}}{\mathop{\mathrm{maximize} }}\,\sum\limits_{m=1}^{N_{\mathrm{cells}}}\sum\limits_{n=1}^{N_{\mathrm{sub}}}{{{w}_{km}}{{r}_{kmn}}}\\
& \text{subject to} \sum\limits_{\begin{smallmatrix}
 n=1 \end{smallmatrix}}^{N_{\mathrm{sub}}}{||{{\bm{f}}_{kmn}}|{{|}_2^{2}}\le {{P}_{m, \max }}},\text{  } m=1,...,N_{\mathrm{cells}}
\end{aligned}
\end{equation}
with $k=\phi(m,n)$. $\bm{\mathcal{F}}:=\left\{{\bm{f}}_{kmn};\text{ }m\in \mathcal{M},\text{ } n\in \mathcal{N}\right\}$ is the set of all beamforming vectors, and $P_{m, \max}$ is the BS transmit power constraint of cell $m$. Because the optimization problem in \eqref{optzm} is non-convex, obtaining the global optimal solution is considerably complex and difficult. Therefore, we focus on a less complex and provably convergent local optimal solution.
\subsection{Review of Second-Order Cone Programming}
There has been substantial development and progress in efficient methods for solving a large class of optimization problems. To employ these algorithms, one should reformulate the optimization problem into a standard form that the algorithms can deal with. Conic programs, i.e., linear programs (LP) with generalized inequalities \cite{Lobo, Boyd}, are subjected to special attention. One such standard conic program for solving convex problems is SOCP, which is of the form
\begin{equation}
\mathrm{SOCP}:\left\{ \begin{matrix}
   \hspace{1mm}\underset{\bm{x}}{\mathrm{maximize }}\ \hspace{3mm}\Re({{ {\bm{a}}}^{\rm{H}}} {\bm{x}}) &  \\
   \hspace{-14mm}\mathrm{subject}\hspace{1mm} \mathrm{to} & \hspace{-15mm}\left[ \begin{matrix}
   {\bm{c}}_{i}^{\rm{H}} {\bm{x}}+{{d}_{i}}  \\
    {\bm{D}}_{i}^{\rm{H}} {\bm{x}}+{{ {\bm{b}}}_{i}}  \\
\end{matrix} \right]{{\succeq }_{\mathcal{K}}}\hspace{1mm}0,\hspace{2mm}i=1,...,U  \\
\end{matrix} \right.
\end{equation}
where $\bm{x}$ is the vector consisting of the optimization variables, whereas  $\bm{a}$, $\bm{b}$, $\bm{c}$, ${d}$, and $\bm{D}_i$ are parameters with appropriate sizes. The notation $\succeq_{\mathcal{K}}$ is used for defining the generalized inequalities as
\begin{equation}
\label{SOCP1}
\left[ (v  \hspace{3mm} {\bm{s}})^{\mathrm{T}}\right]{{\succeq }_{\mathcal{K}}}\hspace{1mm}0\Leftrightarrow || {\bm{s}}|{{|}_{2}}\le v.
\end{equation}
Hyperbolic constraints have an important role in the SOCP formulation of the WSRM objective function and constraints. For hyperbolic equations ${{\bm{z}}^{2}}\le xy,\text{~~}x\ge 0,\text{~~}y\ge 0$ with $\bm{z}\in\mathbb{R}^{1\times n}$ and $x,\hspace{1mm}y\in\mathbb{R}$, the equivalent SOCP is given by \cite{Lobo}
\begin{equation}
\label{SOCP}
{{\bm{z}}^{\mathrm{T}}}{\bm{z}}\le xy,\text{~~}{x}\ge 0,\text{~~}y\ge 0\text{~~}\Leftrightarrow \left\| \begin{matrix}
   2\bm{z}  \\
   x-y  \\
\end{matrix} \right\|_2\le x+y.
\end{equation}
 \begin{figure}
  \centering
   \includegraphics[scale=1.2]{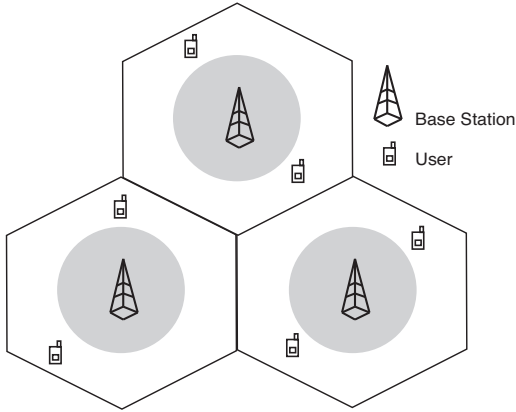}
   \caption{Cell configuration for numerical experiment. }
   \label{cell}
 \end{figure}
\section{Sequential Parametric Convex Approximation for WSRM Problem}
\label{sec:SPCA}
As a first step towards transforming the non-convex WSRM optimization problem in \eqref{optzm} into a standard form that SOCP\footnote{SOCP constraints are convex and can be solved using convex optimization tools such as SeDuMi \cite{Strum}. } is capable of dealing with, we rewrite \eqref{optzm} as  
\begin{equation}
\label{optz}
\begin{aligned}
& \underset{\bm{\mathcal{F}}}{\mathop{\mathrm{maximize} }}\,\sum\limits_{m=1}^{N_{\mathrm{cells}}}{\sum\limits_{\begin{smallmatrix}
 n=1 \\
 k=\phi(m, n)
\end{smallmatrix}}^{N_{\mathrm{sub}}}{{{\alpha }_{kmn}}}{{\log }_{2}}(1+{{\gamma }_{kmn}})}\\
& \text{subject to} \sum\limits_{\begin{smallmatrix}
 n=1 \end{smallmatrix}}^{N_{\mathrm{sub}}}{||{{\bm{f}}_{kmn}}|{{|}_2^{2}}\le {{P}_{m,\max }}},\text{  } m=1,...,N_{\mathrm{cells}}
\end{aligned}
\end{equation}
where ${\alpha }_{kmn}=w_{km},\forall n$. Let $V:=\left\{kmn,\forall m,n\text{  }|\text{  }k=\phi(m,n)\right\}$ and $J=N_{\mathrm{cells}}N_{\mathrm{sub}}$. Therefore, the objective function becomes a function of $J$ variables and can be expressed as
\begin{equation}
\label{optz1}
 \underset{\bm{\mathcal{F}}}{\mathop{\max }}\,\sum\limits_{j=1}^{J}{{{\alpha }_{V_j}}{{\log }_{2}}(1+{{\gamma }_{V_j}})}= \underset{\mathcal{F}}{\mathop{\max }}\,\prod\limits_{j=1}^{J}{{{(1+{{\gamma }_{V_j}})}^{{{\alpha }_{V_j}}}}},
\end{equation}
where $V_j$ refers to the $j$th set in $V$. Replacing ${{{(1+{{\gamma }_{V_j}})}^{{{\alpha }_{V_j}}}}}$ with auxiliary variable ${c}_{V_j}$, we can express the WSRM problem as
\begin{equation}
\label{optz2}
\begin{aligned}
&  \underset{\bm{\mathcal{F}},\hspace{1mm}{{c}_{V_j}}}{\mathop{\mathrm{maximize} }}\,\prod\limits_{j=1}^{J}{{{c}_{V_j}}}\\
& \text{subject to}\\
& \text{~} C1:\sum\limits_{\begin{smallmatrix}
 n=1  \end{smallmatrix}}^{N_{\mathrm{sub}}}{||{{\bm{f}}_{kmn}}|{{|}_2^{2}}\le {{P}_{m,\max }}},\text{  } m=1,...,N_{\mathrm{cells}}\\
&\text{~}C2:c_{V_j}^{{{q}_{V_j}}}\le {{\gamma }_{V_j}}+1,\text{~~}\forall V_j\in V, \text{~}j=1,...,J
\end{aligned}
\end{equation}
with $q_{V_j}=1/{\alpha _{V_j}}$ and the constraints in C2 of \eqref{optz2} active at the optimum. From the definition of ${{\gamma }_{{{V}_{j}}}}$ in \eqref{gamma}, we can rewrite C2 of \eqref{optz2} as
\begin{equation}
c_{{{V}_{j}}}^{{{q}_{{{V}_{j}}}}}-1\le {{\gamma }_{{{V}_{j}}}}=\frac{{{\left| {{\bm{h}}_{{{V}_{j}}}}{{\bm{f}}_{{{V}_{j}}}} \right|}^{2}}}{{\sqrt{1+\sum\limits_{\begin{smallmatrix}{m}'\in \mathcal{M}\backslash m \\
 {k}=\phi(m', n)
 \end{smallmatrix}}{{{\bm{h}}_{km'n}}{{\bm{f}}_{k'm'n}}\bm{f}_{k'm'n}^{\mathrm{H}}\bm{h}_{km'n}^{\mathrm{H}}}}}} ,\nonumber
\end{equation}
which can be equivalently expressed as
\begin{equation}
{{\left( c_{{{V}_{j}}}^{{{q}_{{{V}_{j}}}}}-1 \right)}^{1/2}}\le \frac{{{\bm{h}}_{{{V}_{j}}}}{{\bm{f}}_{{{V}_{j}}}}}{\sqrt{1+\sum\limits_{\begin{smallmatrix}{m}'\in \mathcal{M}\backslash m \\
 {k}=\phi(m', n)
 \end{smallmatrix}}{{{\bm{h}}_{km'n}}{{\bm{f}}_{k'm'n}}\bm{f}_{k'm'n}^{\mathrm{H}}\bm{h}_{km'n}^{\mathrm{H}}}}}.  \nonumber
 \end{equation} 
Further, introducing slack variables $\zeta_{V_j}$ and endorsing $\sqrt{1+\sum\limits_{\begin{smallmatrix}{m}'\in \mathcal{M}\backslash m \\
 {k}=\phi(m', n)
 \end{smallmatrix}}{{{\bm{h}}_{km'n}}{{\bm{f}}_{k'm'n}}\bm{f}_{k'm'n}^{\mathrm{H}}\bm{h}_{km'n}^{\mathrm{H}}}}\le {{\zeta }_{{{V}_{j}}}}$, the resulting WSRM optimization problem from \eqref{optz2} can be rewritten as
\begin{equation}
\label{optz3}
\begin{aligned} 
&  \underset{\bm{\mathcal{F}}, {{c}_{V_j}}, \zeta_{V_j}}{\mathop{\mathrm{maximize} }}\,\prod\limits_{j=1}^{J}{{{c}_{V_j}}}\\
& \text{subject to}\\
&  \text{~} C1:\sum\limits_{\begin{smallmatrix}
 n=1  \end{smallmatrix}}^{N_{\mathrm{sub}}}{||{{\bm{f}}_{kmn}}|{{|}_2^{2}}\le {{P}_{m,\max }}},\text{  } m=1,...,N_{\mathrm{cells}}\\
& \text{~}C2:{{\zeta}_{{{V}_{j}}}}(c_{{{V}_{j}}}^{{{q}_{{{V}_{j}}}}}-1)^{1/2}\le {{\bm{h}}_{V_j}}{{\bm{f}}_{V_j}}\\
&\text{~}C3:\operatorname{\Im}\{{{\bm{h}}_{V_j}}{{\bm{f}}_{V_j}}\}=0\\
&\text{~}C4:\sqrt{1+\sum\limits_{\begin{smallmatrix}{m}'\in \mathcal{M}\backslash m \\
 {k}=\phi(m', n)
 \end{smallmatrix}}{{{\bm{h}}_{km'n}}{{\bm{f}}_{k'm'n}}\bm{f}_{k'm'n}^{\mathrm{H}}\bm{h}_{km'n}^{\mathrm{H}}}}\le {{\zeta }_{{{V}_{j}}}}
\end{aligned}
\end{equation}
The conformity between \eqref{optz2} and \eqref{optz3} can be validated as follows. First, we note that constraining the imaginary part of ${{\bm{h}}_{V_j}}{{\bm{f}}_{V_j}}$ to zero in C2 of \eqref{optz3} does not affect the optimality\footnote{For any $\pi$, we have $|\bm{h}_{V_j}\bm{f}_{V_j}|^2=|\bm{h}_{V_j}\bm{f}_{V_j}e^{j\pi}|^2$. Therefore, choosing $\pi$ such that $\Im\{{{\bm{h}}_{V_j}}{{\bm{f}}_{V_j}}\}=0$ does not affect optimality.} of \eqref{optz2}. Second, we can clarify that at the optimum, all the constraints in C4 of \eqref{optz3} uphold the equality. For instance, let us presume that the $V_j$th constraint in C4 is not active. Let $\bar{{{\zeta }_{{{V}_{j}}}}}\triangleq {{\zeta }_{{{V}_{j}}}}/\beta$ and ${{\bar{c}}_{{{V}_{j}}}}\triangleq {{\left\{ 1+\left( c_{{{V}_{j}}}^{{{q}_{{{V}_{j}}}}}-1 \right){{\beta }^{2}} \right\}}^{1/{{q}_{{{V}_{j}}}}}}$, where $\beta$ is a positive scaling parameter. Choosing $\beta>1$, the constraints in C2 and C4 of \eqref{optz3} become active together if we substitute $\left( {{\zeta }_{{{V}_{j}}}},{{c}_{{{V}_{J}}}} \right)$ with $\left( {{{\bar{\zeta }}}_{{{V}_{j}}}},{{{\bar{c}}}_{{{V}_{J}}}} \right)$.
However, the impact of such a substitution is factually a larger objective because ${{\bar{c}}_{{{V}_{J}}}}>{{c}_{{{V}_{j}}}}$ for $\beta >1$, which is a contradiction with the fact that an optimal solution is obtained.

Now, we clearly notice that per-cell transmit power constraints C1, resulting SINR constraints C2, and constraint C3 are LPs with generalized equalities/inequalities that can directly be expressed as SOCP. Because these constraints are already in convex form, they require no approximation. In order to express C1 as SOCP, let $\bm{\mathcal{F}}_m$ be the set of beamformers for cell $m$. By making use of operator $\mathrm{vec}(\cdot)$, we can reformulate C1 as $|| \mathrm{vec}\left( {{\bm{\mathcal{F}}}_{m}} \right)|{{|}_{2}}\le \sqrt{{{P}_{m,\max }}}$, for which the equivalent SOCP according to \eqref{SOCP1} is given by
\begin{equation}
\label{eq44}
\left[ \begin{matrix}
 \sqrt{{{P}_{m,\max }}}    \\ 
 \mathrm{vec}\left( {{\bm{\mathcal{F}}}_{m}} \right)  \\
\end{matrix} \right]{{\succeq }_{\mathcal{K}}}\hspace{1mm}0.
\end{equation}
To express C4 of \eqref{optz3} as an SOCP, let $\bm{H}_{\operatorname{int}}\in\mathbb{C}^{{(N_{\mathrm{cells}}-1)\times N_\mathrm{Tx}}} $ and $\bm{F}_{\operatorname{int}}\in\mathbb{C}^{N_\mathrm{Tx} \times(N_{\mathrm{cells}}-1)}$ be the collected channel and beamforming matrices, respectively, containing the channels from all interfering BSs and beamforming vectors corresponding to constraint C4 of \eqref{optz3}. As a result, we can write constraint C4 as ${{\left\| {{\left[ \begin{matrix}1 & \mathrm{diag}\left( {{\bm{H}}_{\operatorname{int}}}{{\bm{F}}_{\operatorname{int}}} \right)  \\ \end{matrix} \right]}^{\mathrm{T}}} \right\|}_{2}}\le {{\zeta }_{{{V}_{j}}}}$, which is equivalent to an SOCP
\begin{equation}
\label{equat33}
\left[ \begin{matrix}
   {{\zeta }_{{{V}_{j}}}}  \\
   {{\left[ \begin{matrix}
   1 & \mathrm{diag}\left( {{\bm{H}}_{\operatorname{int}}}{{\bm{F}}_{\operatorname{int}}} \right)  \\
\end{matrix} \right]}^{\mathrm{T}}}  \\
\end{matrix} \right]{{\succeq }_{\mathcal{K}}}\hspace{1mm}0.
\end{equation}
We are now left with non-convex constraint C2 of the optimization problem in \eqref{optz3}. We exploit the SPCA approach to approximate C2 as convex. To initiate the convex approximation, we rewrite C2 with a suitable change of variables as
\begin{equation}
\label{eq5}
p_{{{V}_{j}}}^{1/2}{{\zeta }_{{{V}_{j}}}}\le {{\bm{h}}_{V_j}}{{\bm{f}}_{V_j}},\hspace{1mm}\forall {{V}_{j}}\in V
\end{equation}
\begin{equation}
\label{eq6}
c_{{{V}_{j}}}^{{{q}_{{{V}_{j}}}}}\le {{p}_{{{V}_{j}}}}+1.
\end{equation}
Both \eqref{eq5} and \eqref{eq6} are still non-convex, however, and this formulation leads us to employ the SPCA technique. First, we focus on the convex approximation of \eqref{eq5}. Defining $\mathcal{Q}({{\zeta}_{{{V}_{j}}}},{{p}_{{{V}_{j}}}})=p_{{{V}_{j}}}^{1/2}{{\zeta }_{{{V}_{j}}}}$ with $p_{{{V}_{j}}},\text{ }{{\zeta }_{{{V}_{j}}}}\ge0$, we approximate $\mathcal{Q}({{\zeta}_{{{V}_{j}}}},{{p}_{{{V}_{j}}}})$ with its convex upper estimate function $U({{\zeta }_{{{V}_{j}}}},{{p}_{{{V}_{j}}}},{{\theta }_{{{V}_{j}}}})$ according to \cite{Beck} as
\begin{equation}
\label{eq1}
U({{\zeta }_{{{V}_{j}}}},{{p}_{{{V}_{j}}}},{{\theta }_{{{V}_{j}}}})\triangleq \frac{1}{2}\left( \frac{{{p}_{{{V}_{j}}}}}{{{\theta }_{{{V}_{j}}}}}+{{\theta }_{{{V}_{j}}}}\zeta _{{{V}_{j}}}^{2} \right).
\end{equation}
Consequently, $\mathcal{Q}({{\zeta }_{{{V}_{j}}}},{{p}_{{{V}_{j}}}})\le U({{\zeta }_{{{V}_{j}}}},{{p}_{{{V}_{j}}}},{{\theta }_{{{V}_{j}}}})$, $\forall {\theta }_{{V}_{j}}\ge 0$. At the optimum, $\mathcal{Q}({{\zeta}_{{{V}_{j}}}},{{p}_{{{V}_{j}}}})=U({{\zeta}_{{{V}_{j}}}},{{p}_{{{V}_{j}}}},{{\theta}_{{{V}_{j}}}})$ with ${{\theta }_{{{V}_{j}}}}=\sqrt{{{p}_{{{V}_{j}}}}}/{{\zeta }_{{{V}_{j}}}}$. Using the successive approximation method, this optimal point is approached in an iterative way by intuitively updating the variables until the KKT points of \eqref{optz3} are obtained. This convex over-estimation of $\mathcal{Q}({{\zeta}_{{{V}_{j}}}},{{p}_{{{V}_{j}}}})$ allows us to express equation \eqref{eq5} as a hyperbolic constraint as ${{\left\| {{\left[ {{\zeta }_{{{V}_{j}}}}\sqrt{\frac{\theta _{{{V}_{j}}}}{2}}\text{~~}({{\bm{h}}_{V_j}}{{\bm{f}}_{V_j}}-\frac{{{p}_{{{V}_{j}}}}}{2\theta _{{{V}_{j}}}}-1)] \right]}^{\mathrm{T}}} \right\|}_{2}}\le ({{\bm{h}}_{V_j}}{{\bm{f}}_{V_j}}-\frac{{{p}_{{{V}_{j}}}}}{2\theta _{{{V}_{j}}}}+1)$,
and the corresponding SOCP representation is given by
\begin{equation}
\label{equat22}
\left[ \begin{matrix}
   {{\bm{h}}_{{{V}_{j}}}}{{\bm{f}}_{{{V}_{j}}}}-\frac{{{p}_{{{V}_{j}}}}}{2\theta _{{{V}_{j}}}}+1  \\
   {{\left[ {{\zeta }_{{{V}_{j}}}}\sqrt{\frac{\theta _{{{V}_{j}}}}{2}}\hspace{4mm}({{\bm{h}}_{{{V}_{j}}}}{{\bm{f}}_{{{V}_{j}}}}-\frac{{{p}_{{{V}_{j}}}}}{2\theta _{{{V}_{j}}}}-1) \right]}^{\mathrm{T}}}  \\
\end{matrix} \right]{{\succeq }_{\mathcal{K}}}\hspace{1mm}0.
\end{equation}
Now, let us turn our focus onto the convex approximation of \eqref{eq6}. To arrive at an SOCP, we scale all $q_{V_j}$ such that $q_{V_j}<1$ so as to make function $c_{{{V}_{j}}}^{{{q}_{{{V}_{j}}}}}$ concave. Function $c_{{{V}_{j}}}^{{{q}_{{{V}_{j}}}}}$ is differentiable. For differentiable concave function $\mathcal{V}$ with $(\forall x,y\in\text{domain }(\mathcal{V}))$, the first-order concavity condition says that the gradient line is the global over-estimator of the function \cite{Lobo}. Function $\mathcal{V}(x)+{{\nabla}_{x}}\mathcal{V}{{(x)}^{T}}(y-x)$ is defined as a first-order linear approximation to the function at $x$, where ${{\left({{\nabla}_{x}}\mathcal{V}(x)\right)}_{i}}=\frac{\partial \mathcal{V}(x)}{\partial {{x}_{i}}}$. Equivalently, we approximate $c_{V_j}^{{{q}_{V_j}}}$ with its concave over-estimator as follows
\begin{equation}
\label{equat1}
\begin{aligned}
& c_{{{V}_{j}}}^{{{q}_{{{V}_{j}}}}}-c_{{{V}_{j}},i}^{{{q}_{{{V}_{j}}}}}\le {{q}_{{{V}_{j}}}}c_{{{V}_{j}},i}^{{{q}_{{{V}_{j}}}}-1}({{c}_{{{V}_{j}}}}-{{c}_{{{V}_{j}},i}})\\
&  {{p}_{{{V}_{j}}}}\ge {{q}_{{{V}_{j}}}}c_{{{V}_{j}},i}^{{{q}_{{{V}_{j}}}}-1}({{c}_{{{V}_{j}}}}-{{c}_{{{V}_{j}},i}})+c_{{{V}_{j}},i}^{{{q}_{{{V}_{j}}}}}-1 \text{ }(\text{with \eqref{eq6}})\\
\end{aligned}
\end{equation}
and iteratively solve \eqref{equat1} in parallel with \eqref{equat22} until convergence. Specifically, it is the linearization of $c_{{{V}_{j}}}^{{{q}_{{{V}_{j}}}}}$ around point ${{c}_{{{V}_{j}},i}}$, where ${{c}_{{{V}_{j}},i}}$ is the value of ${{c}_{{{V}_{j}}}}$ at the $i$th iteration. Both \eqref{equat22} and \eqref{equat1} are non-decreasing; however, they are upper-bounded by the per-BS transmit power constraints.

Finally, we focus on reformulating the objective function of the WSRM problem in a convex solvable form. There are two possible ways to express the objective function of \eqref{optz3} such that the objective function can be handled by the existing convex solver, as described here.

\textbf{Method 1:} The geometric mean (GM) of the optimization 
variables, $\chi=({{c}_{V_1}}{{c}_{V_2}}...{{c}_{V_{J}}})^{1/{J}}$, is concave when ${{c}_{V_j}}\succeq 0,\forall V_j$. Maximizing the GM of the optimization variables will yield the same weighted sum-rate as maximizing the product of the optimization variables as long as the variables are nonnegative affine\cite{Lobo}; hence, we can rewrite the objective function as
\begin{equation}
\label{eqGM}
\underset{\bm{\mathcal{F}},{{c}_{V_j}},\zeta_{V_j}}{\mathop{\mathrm{maximize} }}\,\prod\limits_{j=1}^{J}{{{c}_{V_j}}}:\Leftrightarrow \underset{\bm{\mathcal{F}},{{c}_{V_j}},\zeta_{V_j}}{\mathop{\mathrm{maximize} }}\,\chi.
\end{equation}
Using the CVX\cite{CVX} solver with SeduMi, a disciplined convex programming, we can directly use the GM of the optimization variables as an objective function.

\textbf{Method 2:} The second method is based on transformation of the product of the optimization variables into hyperbolic constraints, which also admit SOCP representation. Thus, we need to reformulate the problem by introducing new variables and by incorporating hyperbolic constraints.
Let $\psi$ be the set of new variables. During the transformation process, variables in $\psi$ are assigned values at $\log_2{J}$ stages. To simplify analysis, let $J=2^{u}$, where $u$ is a real positive quantity. The transformation procedure is provided below.
%
\vspace{-6mm}
\begin{center}
\line(1,0){240}
\end{center}
\vspace{-3mm}
\textbf{Procedure 1:} \text{For hyperbolic constraints formulation}
\vspace{-5mm}
\begin{center}
\nointerlineskip
\line(1,0){240}
\end{center}
\vspace{-3mm}
${\textbf{Initialize:  }} \psi _{j}^{u}={{c}_{{{V}_{j}}}},\text{  }j=1,...,J\text{  and  } u={\log}_{2}(J)$\\
$\textbf{for  }l=u,u-1,...,1$\\
$ {{\left( \psi _{i}^{l-1} \right)}^{2}}\le \psi _{2i-1}^{l}\psi _{2i}^{l},\text{  }i=1,...,{{2}^{l-1}}$\\
\textbf{end}
\vspace{-6mm}
\begin{center}
\nointerlineskip
\line(1,0){240}
\end{center}

At the last stage of the hyperbolic constraints transformation process, the objective function emerges to be a one-variable optimization problem defined as $\psi_1^0=\psi^0$. Finally, applying \eqref{SOCP} yields the SOCP formulations for $2^u-1$ hyperbolic equations of \textbf{Method 2}. 
It is worth noting that this algorithm is inspired by \cite{Beck, Chris, Tran} and is similar to \cite{Tran}, which proposes the SPCA-based algorithm for multicell MU-MISO networks. However, we formulate and propose the SPCA-based algorithm with a GM approach for multicell OFDMA networks and resolve two practical limiting factors related to the algorithm implementation, which are not addressed in \cite{Tran} to make the algorithm more general, especially when the problem size is comparatively larger.
\vspace{-6mm}
\begin{center}
\line(1,0){240}
\end{center}
\vspace{-3mm}
\textbf{Algorithm: } SOCP-based SPCA-WSRM
\vspace{-5mm}
\begin{center}
\nointerlineskip
\line(1,0){240}
\end{center}
\textbf{1. Initialization:  } $N_{\rm{iter}}$,\hspace{1mm}$i=0$, \hspace{1mm}$(\theta_{V_j}^{i}, \hspace{1mm}c_{V_j,i})$, \\
\textbf{2. Repeat  }\\
\textbf{3. Solve the following  }\\
\text{~~~~~~~~}$\underset{\bm{\mathcal{F}},{{c}_{V_j}},\zeta_{V_j}}{\mathop{\mathrm{maximize} }}\,\chi\hspace{1mm} \text{ (if \textbf{Method 1} is used) or}$\\
\text{~~~~~~~~}$\underset{\bm{\mathcal{F}},{{c}_{V_j}},\zeta_{V_j}, v_{V_j}, \psi_{V_j}}{\mathop{\mathrm{maximize} }}\,\psi^0 \hspace{1mm}\text{ (if \textbf{Method 2} is used)}$\\
\text{~~~~~~~~}\text{subject to }\\
\text{~~~~~~~~~~~~~~}C1: {\textbf{Procedure 1 }} {\text{with \eqref{SOCP} (if \textbf{Method 2} is used)}}.\\
\text{~~~~~~~~~~~~~~}C2: $\left[ \begin{matrix}
  \sqrt{{{P}_{m,\max }}} \\
\mathrm{vec}\left( {{\bm{\mathcal{F}}}_{m}} \right) \\
\end{matrix} \right]{{\succeq }_{\mathcal{K}}}\hspace{1mm}0,\text{  } m=1,...,N_{\mathrm{cells}}$.\\
\text{~~~~~~~~~~~~~~}C3: $\left[ \begin{matrix}
   {{\bm{h}}_{{{V}_{j}}}}{{\bm{f}}_{{{V}_{j}}}}-\frac{{{p}_{{{V}_{j}}}}}{2\theta _{{{V}_{j}}}^{i}}+1  \\
   {{\left[ {{\zeta }_{{{V}_{j}}}}\sqrt{\frac{\theta _{{{V}_{j}}}^{i}}{2}}\hspace{4mm}({{\bm{h}}_{{{V}_{j}}}}{{\bm{f}}_{{{V}_{j}}}}-\frac{{{p}_{{{V}_{j}}}}}{2\theta _{{{V}_{j}}}^{i}}-1) \right]}^{\mathrm{T}}}  \\
\end{matrix} \right]{{\succeq }_{\mathcal{K}}}\hspace{1mm}0$.\\
\text{~~~~~~~~~~~~~~}C4: $\operatorname{\Im}\{{{\bm{h}}_{V_j}}{{\bm{f}}_{V_j}}\}=0$.\\
\text{~~~~~~~~~~~~~~}C5: ${{p}_{{{V}_{j}}}}\ge {{q}_{{{V}_{j}}}}c_{{{V}_{j}},i}^{{{q}_{{{V}_{j}}}}-1}({{c}_{{{V}_{j}}}}-{{c}_{{{V}_{j}},i}})+c_{{{V}_{j}},i}^{{{q}_{{{V}_{j}}}}}-1$.\\
\text{~~~~~~~~~~~~~~}C6: $\left[ \begin{matrix}
   {{\zeta }_{{{V}_{j}}}}  \\
   {{\left[ \begin{matrix}
   1 & \mathrm{diag}\left( {{\bm{H}}_{\operatorname{int}}}{{\bm{F}}_{\operatorname{int}}} \right)  \\
\end{matrix} \right]}^{\mathrm{T}}}  \\
\end{matrix} \right]{{\succeq }_{\mathcal{K}}}\hspace{1mm}0$.\\
\text{~~~~~~~~~~~~~~}C7: $\psi_{V_j}\ge0, c_{V_j}\ge0,\text{~  } \text{ implicit constraints }$.\\
\textbf{4.}$\text{ Denote }(c_{V_j,i+1}, \zeta_{V_j}^{i+1}, p_{V_j}^{i+1})= \text{optimal values at step 3.}$\\
\textbf{5.} $\theta _{{V_{j}}}^{i+1}=\sqrt{v_{V_j}^{i+1}}/\zeta_{V_j}^{i+1},\text{ }\text{ }i=i+1$\\
\textbf{6.} $\textbf{until }\text{convergence or } i=N_{\rm{iter}}$\\
\vspace{-6mm}
\begin{center}
\nointerlineskip
\line(1,0){240}
\end{center}
In the iterative optimization process, the initial ${{\theta }_{{{V}_{j}}}}$s are crucial to feasibility and convergence. We have noticed that in most cases, the randomly generated ${{\theta }_{{{V}_{j}}}}$s lead to infeasible solutions in the first iteration. To make sure that the
algorithm is feasible on the first step, we follow the steps in \textbf{Procedure 2} to find good initial ${{\theta }_{{{V}_{j}}}}$s.

The other numerical issue that is not addressed in \cite{Tran} is the situation when one or more of the
$v_{V_j}$s become zero, i.e., no power in that or those particular subcarriers of the corresponding cells. It is normal for some of the subcarriers to not have any power because of limited BS power if we recall the mechanism of a water-filling algorithm. However, when such a situation arises, we have noticed numerical instability. We encounter the problem of dividing by zero because we need to calculate $1/\theta_{V_j}$. In order to avoid this situation, we slightly modify the imposed constraints on $p_{V_j}$ such as $p_{V_j}\ge \varepsilon$ (e.g., $\varepsilon$=0.0001) so as to bypass the numerical problem.
This modification yields a solution that is close to the original one without encountering numerical instability.
%
\vspace{-10mm}
\begin{center}
\line(1,0){240}
\end{center}
\vspace{-3mm}
$\textbf{Procedure 2:} \text{ For generating initial   } {{\theta }_{{{V}_{j}}}}\rm{s}$
\vspace{-5mm}
\begin{center}
\nointerlineskip
\line(1,0){240}
\end{center}
\vspace{-3mm}
${\textbf{Step 1: }}\text{Generate channel-matched beamformers such that}$\\
$\text{the per-BS power constraint is satisfied for all cells, } \text{i.e., }{{\bm{f}}_{kmn}}=\sqrt{{{{P}_{m,\max }}}/{N}}({{{\bm{h}}_{kmn}}}/{||{{\bm{h}}_{kmn}}||_2}) \text{ with }k=\phi(m,n).$\\
${\textbf{Step 2: }}\text{ Use C4 of \eqref{optz3} to find }{\zeta }_{{V}_{j}}\text{ by substituting the ineq- }$\\
$\text{uality with an equality.}$\\
${\textbf{Step 3: }}\text{Calculate }c_{V_{j},0}\text{ from  C2 of \eqref{optz3} manipulating the ab- }$\\
$\text{solute value of } {{\bm{h}}_{V_j}}{{\bm{f}}_{V_j}}.$\\
$ {\textbf{Step 4: }}\text{Find }p_{V_j}\text{ using \eqref{eq5}}.\text{ Finally, the initial value of }{{\theta }_{{{V}_{j}}}}$\\
$\text{is obtained as } {\theta }_{V_j}^0=\sqrt{{{p}_{{{V}_{j}}}}}/{{\zeta }_{{{V}_{j}}}}.$\\
\vspace{-8mm}
\begin{center}
\nointerlineskip
\line(1,0){240}
\end{center}
\section{Power Optimization with SINR Constraints}
In this section, we consider the power optimization issue of multicell downlink beamforming subject to individual SINR constraints of the subcarriers per cell. Generally, the SINR constraints are determined based on the signaling scheme and the target bit error rate (BER) requirements of the system. The SINRs are dependent on the transmission power and the choice of beamformers. In our study, minimization is performed over the beamformers because the power is absorbed by them. 
\begin{equation}
\label{sinr1}
\begin{aligned} 
& \underset{\bm{\mathcal{F}}}{\mathrm{minimize }}\,\sum\limits_{m=1}^{N_{\mathrm{cells}}}\sum\limits_{n=1}^{N_{\mathrm{sub}}}{||{{\bm{f}}_{kmn}}||_{2}^{2}}\\
& \text{subject to}\text{~}{{\gamma }_{mkn}}\ge {{\Gamma }_{mkn}},\forall m,n \text{ and } k=\phi(m,n),\\
\end{aligned}
\end{equation}
where ${{\Gamma }_{mkn}}$ is the target SINR for subcarrier $n$ of cell $m$, which is assigned to user $k=\phi(m,n)$. From a network operator's perspective, this strategy is interesting because it optimizes the power efficiency of the system through minimization of the inter-carrier interference.
When devising an algorithmic solution to the optimization problem in \eqref{sinr1}, we are required to clarify the possibility that the SINR constraints may be infeasible. Throughout this analysis, we assume that the SINR targets are feasible. SINR feasibility is guaranteed because the SINR constraints we employ in this analysis are actually taken from previous WSRM optimization. 

Per-subcarrier SINR constraints in \eqref{sinr1} may appear non-convex. For a single-cell downlink beamforming case, the authors of \cite{Wiesel} showed that the SINR constraints of this type can be formulated as SOCP constraints. This important observation motivates us to solve \eqref{sinr1} via an SOCP-based convex optimization program. Using the SINR definition in \eqref{gamma}, we can rewrite the SINR constraints of \eqref{sinr1} as
\begin{equation}
\label{sinr2}
\frac{1}{{{\Gamma }_{kmn}}}|{{\bm{h}}_{kmn}}{{\bm{f}}_{kmn}}{{|}^{2}}\ge \left\| {{\left[ \begin{matrix}
   1 & \mathrm{diag}\left( {{\bm{H}}_{\operatorname{int}}}{{\bm{F}}_{\operatorname{int}}} \right)  \\
\end{matrix} \right]}^{\mathrm{T}}} \right\|_{2}^{2}.
\end{equation}
In the previous section, as we restricted ourselves to the beamformers in which ${{\bm{h}}_{kmn}}{{\bm{f}}_{kmn}}\ge 0,\hspace{1mm}\forall m,n$, \eqref{sinr2} results in
\begin{equation}
\sqrt{\frac{1}{{{\Gamma }_{kmn}}}}{{\bm{h}}_{kmn}}{{\bm{f}}_{kmn}}\ge {{\left\| {{\left[ \begin{matrix}
   1 & \mathrm{diag}\left( {{\bm{H}}_{\operatorname{int}}}{{\bm{F}}_{\operatorname{int}}} \right)  \\
\end{matrix} \right]}^{\mathrm{T}}} \right\|}_{2}},
\end{equation}
which can be formulated as SOCP
\begin{equation}
\left[ \begin{matrix}
   \sqrt{\frac{1}{{{\Gamma }_{kmn}}}}{{\bm{h}}_{kmn}}{{\bm{f}}_{kmn}}  \\
   {{\left[ \begin{matrix}
   1 & \mathrm{diag}\left( {{\bm{H}}_{\operatorname{int}}}{{\bm{F}}_{\operatorname{int}}} \right)  \\
\end{matrix} \right]}^{\mathrm{T}}}  \\
\end{matrix} \right]{{\succeq }_{\mathcal{K}}}\hspace{1mm}0.
\end{equation}
The power optimization problem can also be defined under per-cell maximum transmit power constraints, as we considered in the previous section. Denoting $\xi= \sqrt{{{P}_{m,\max }}}$, the SOCP-based optimization for problem \eqref{sinr1} becomes
\begin{equation}
\label{optz6}
\begin{aligned} 
&  \underset{\bm{\mathcal{F}},\text{~}\xi}{\mathrm{minimize }}\,\text{~}\xi\\
& \text{subject to}\text{~~} C1:\operatorname{\Im}\{{{\bm{h}}_{kmn}}{{\bm{f}}_{kmn}}\}=0\\
&\text{~~~~~~~~~~~~~~~~~}C2:\left[ \begin{matrix}
   \sqrt{\frac{1}{{{\Gamma }_{kmn}}}}{{\bm{h}}_{kmn}}{{\bm{f}}_{kmn}}  \\
   {{\left[ \begin{matrix}
   1 & \mathrm{diag}\left( {{\bm{H}}_{\operatorname{int}}}{{\bm{F}}_{\operatorname{int}}} \right)  \\
\end{matrix} \right]}^{\mathrm{T}}}  \\
\end{matrix} \right]{{\succeq }_{\mathcal{K}}}\hspace{1mm}0\\
& \text{~~~~~~~~~~~~~~~~~}\text{}C3:\left[ \begin{matrix}
  \mathrm{vec}\left( {{\mathcal{F}}_{m}} \right)  \\
   \xi  \\
\end{matrix} \right]{{\succeq }_{\mathcal{K}}}\hspace{1mm}0\\
\end{aligned}
\end{equation}
Relaxing constraint C3 of \eqref{optz6} means that there is no per-cell transmit power constraint. However, as long as the SINR constraints are feasible under the availability of per-cell transmit power, constraint C3 has no significant impact. 
\section{Numerical Results}
The performance of the proposed algorithm is analyzed on a cellular network with three coordinated BSs, two users per cell, and one cell frequency reuse factor. The distance between adjacent BSs is 1000 m. The users are uniformly distributed around their own BS within a circular annulus with external and internal radii of 1000 m and 500 m, respectively, as shown in Fig.~\ref{cell}. Similar to \cite{Venturino}, frequency-selective channel coefficients over 64 subcarriers are modeled as ${{\bm{h}}_{kmn}}={{\left(200\frac{1}{{{l}_{km}}}\right)}^{3.5}}{{\Phi }_{kmn}}{{\Lambda }_{kmn}},$
where ${l}_{km}$ is the distance between BS $m$ and user $k$. The value of $10{{\log }_{10}}({{\Phi }_{kmn}})$ is distributed as $\mathcal{RN}(0,8)$, accounting for log-normal shadowing, and ${{\Lambda }_{kmn}}\sim\mathcal{CN}(0,1)$ accounts for Rayleigh fading. All BSs are subjected to equal transmit power constraints, i.e., ${{P}_{m,\max }}={{P}_{\max }},\forall m$. We also consider that perfect channel state information (CSI) is available both at the BSs and users. The initial user assignment is performed randomly. We consider $N_\mathrm{Tx}=2$ and use the CVX \cite{CVX} package for specifying and solving convex programs. For all cases, the maximum number of iterations, $N_{\rm{iter}}$, is 20.
\begin{figure}
  \centering
   \includegraphics[scale=1.2]{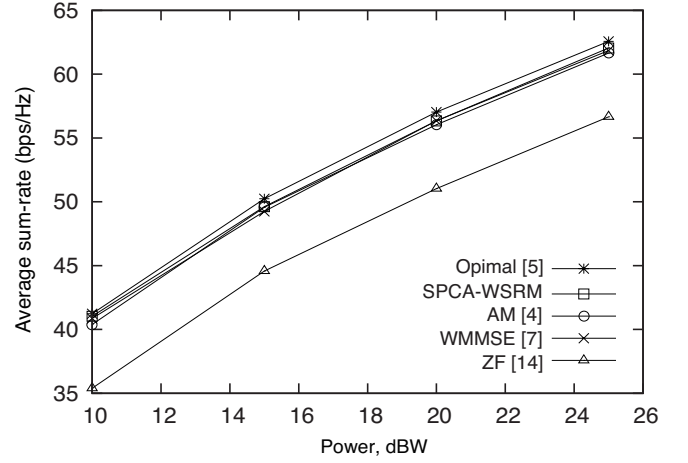}
   \caption{Average sum-rate performances for different WSRM algorithms.  }
   \label{BER3}
 \end{figure}
 
 We compare the average sum-rate ($w_{km}$=1, $\forall m, k$) performances for various precoding strategies as a function of per-BS transmit power in Fig.~\ref{BER3}. The iterative procedure for SPCA-WSRM and other iterative methods, such as AM \cite{Wang} and WMMSE \cite{Christensen}, stop when the increase in objective value between two successive iterations is $\le\epsilon \hspace{1mm} (=0.01)$. We see that the suboptimal solutions achieved by the SPCA-WSRM algorithm and other techniques such as AM and WMMSE are indeed very close to the optimal precoding performance achieved by the branch-and-bound (BB) method \cite{Joshi}. The gap tolerance between the upper and lower bound of the BB method is set to 0.01. Though the sum-rate achieved by SPCA-WSRM, AM, and WMMSE are very close, the convergence rates of the optimal BB, AM, and WMMSE algorithms are significantly slower. 
 
From the point of view of complexity, though the SPCA-WSRM algorithm consists of a large set of constraints, they are all SOCP constraints, i.e., convex. For the $m$th cell, the computational complexity per iteration is $O\left( \sum\limits_{k=1}^{N_{\mathrm{users/cell}}}{ D({{\mathcal{S}}_{km}}) {{N}_{\rm{Tx}}}} \right)$, where $D({{\mathcal{S}}_{km}})$ defines the cardinality of ${{\mathcal{S}}_{km}}$. Moreover, the SOCP in each iteration of SPCA-WSRM is a sparse SOCP, i.e., a large number of zeros appear in the KKT matrix. The sparsity increases along with the iterative procedure. Because operations pertaining to sparse matrix structures and algorithms are faster, our proposed SPCA-SOCP method converges within a small number of iterations. The most significant aspect of this approach is that it is general enough to apply to a  variety of problems relating to SINR optimization (not necessarily some-rate).

The computational complexities of both AM and WMMSE per iteration are almost the same and depend on the problem size. A closed-form expression can be found per iteration for the case of the sum-power constraint for any $m$th cell. The WMMSE admits an water-filling solution per iteration if we consider sum-power constraint, whereas if we consider a special case of per-antenna power constraints in each cell, the complexity of WMMSE is equal to that of the SPCA-WSRM algorithm. However, the convergence speed of WMMSE is slow since it is based on alternating optimization concept. Although the BB algorithm is guaranteed to achieve an optimal solution, it is very computationally expensive. With a gap of 0.01, the BB method converges to a sum-rate, which is close enough to the optimal sum-rate. However, it takes approximately 900 iterations to converge to this sum-rate. 

 \begin{figure}
  \centering
   \includegraphics[scale=1.2]{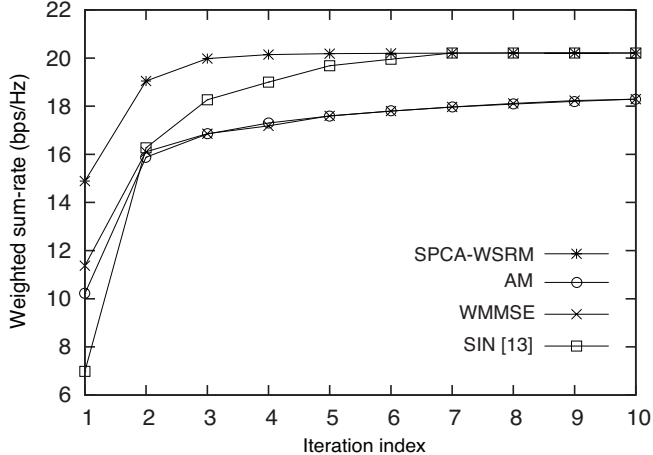}
   \caption{Convergence rate comparison for different WSRM algorithms. }
   \label{BER1}
 \end{figure}
 \begin{figure}
  \centering
   \includegraphics[scale=1.2]{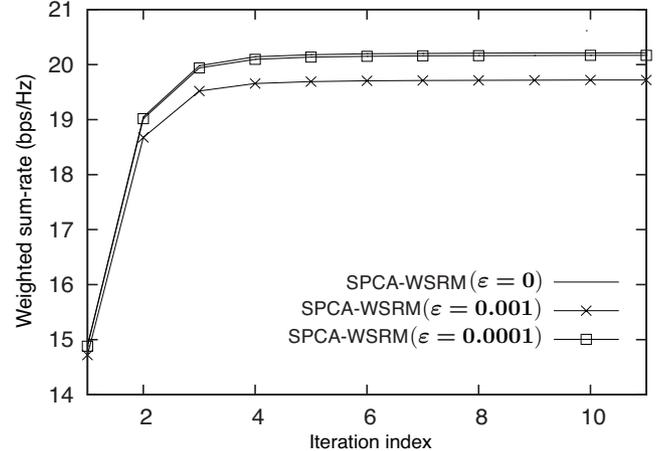}
   \caption{Performance of SPCA-WSRM with modified constraint on $p_{V_j}$.}
   \label{BER2}
 \end{figure}
 \begin{figure*}
  \centering
   \includegraphics{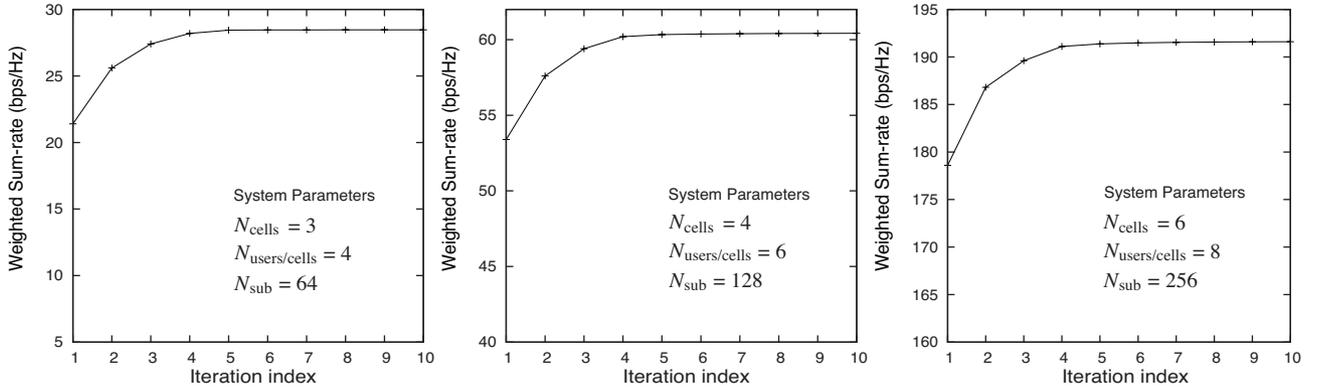}
   \caption{SPCA-WSRM convergence performances for different system parameters.}
   \label{Rev}
\end{figure*}

In Fig.~\ref{BER1}, we compare the WSRs achieved by the proposed SPCA-WSRM method and other iterative algorithms as a function of the number of iterations required to acquire steady output for a random channel realization. The maximum transmit power limit for all the BSs is set to 20 dBW, and the user weights are chosen to lie within the range from 0.10 to 0.60. We observe that the SPCA-WSRM algorithm converges within a few iterations, whereas the AM and WMMSE algorithms are still far away from the convergence level of SPCA-WSRM. This incident may be attributed to the fact that AM-WSRM algorithm alternates between a closed-form posterior conditional probability update and updating the beamforming vectors, whereas the WMMSE algorithm relies on the relationship between the mutual information and minimum mean-square error (MMSE) and alternates between the updating of receive and transmit beamformers. As a result, comparatively slower convergences are observed. However, good initial values for the variables involved in WMMSE accelerate the convergence rate. The successive interference nulling (SIN) algorithm \cite{Chris} is based on solving determinant-maximization (MAXDET) programs. So the complexity is very high when we have large number of antennas at BS. Though it exhibits convergence performance close to the SPCA-WSRM, the per-iteration running time is approximately 5 to 6 times higher.  

Fig.~\ref{BER2} compares the WSR performances of SPCA-based optimization for different values of $\varepsilon$ such that $p_{V_j}\ge \varepsilon$. For all the cases, we generate the initial values of ${{\theta }_{{{V}_{j}}}}$s using \textbf{Procedure 2}. We notice that the modified constraints do not significantly affect optimality. It is obvious that the larger the value of $\varepsilon$, the larger the performance gap between the curves evolves. We do not depict any explicit capacity plot for \textbf{Method 1} and \textbf{Method 2} because both of them exhibit identical WSR performance. Although both methods produce equivalent WSR, the per-iteration running time for \textbf{Method 1} is slightly longer than \textbf{Method 2}. We have also observed that the SPCA-WSRM algorithm along with \textbf{Procedure 2} provides a feasible solution to the optimization problem all the times. 
%

In Fig.~\ref{Rev}, we plot the convergence performance curves of our proposed WSRM method for different system parameters. The curves are obtained for three different cases: case 1 with $\{N_{\rm{cells}}=3, N_{\rm{users/cell}}=4, N_{\rm{sub}}=64 \}$, case 2 with $\{N_{\rm{cells}}=4, N_{\rm{users/cell}}=6, N_{\rm{sub}}=128 \}$, and case 3 with $\{N_{\rm{cells}}=6, N_{\rm{users/cell}}=8, N_{\rm{sub}}=256 \}$. Because the SPCA-WSRM method is a linear-time method, the complexity per cell per iteration increases linearly with increases in $N_{\rm{Tx}}$, $N_{\rm{users/cell}}$, and $N_{\rm{sub}}$. As a result, the per-iteration running time increases. However, the iterative convergence behaviors remain almost the same.
 \begin{figure}
  \centering
   \includegraphics[scale=1.2]{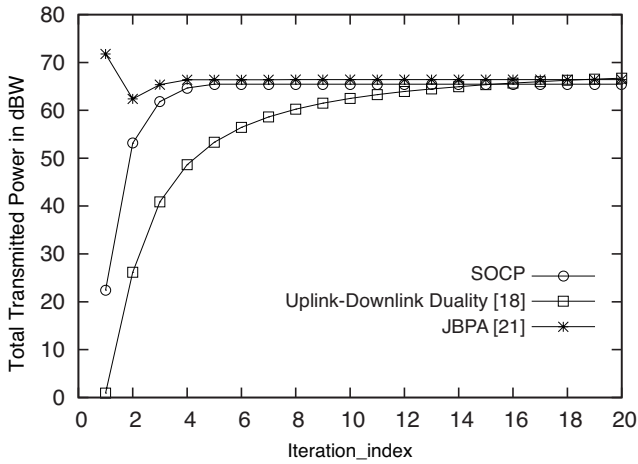}
   \caption{Total transmitted power versus iteration indexes for joint optimization of a coordinated beamforming system.}
   \label{power}
 \end{figure}

Finally, in Fig.~\ref{power}, we compare the convergence behaviors for different power optimization algorithms. Though any initial power vector sequence will converge to the optimal power allocations in \cite{Rashid2}, they have a rapid power fluctuation impact at the very first iteration if the initial power allocations are not chosen appropriately. For the current analysis, the initial power values are chosen to be 0.006 W over all the subcarriers. The rate optimization based on uplink-downlink duality has a comparatively slower convergence rate due to alternate updating between the uplink beamformer and downlink transmit power. Though all these iterative methods converge to the global optimal solution, they exhibit different convergence rates. It is clear that SOCP-based optimization has a smooth and faster convergence as compared to the other iterative methods evaluated. For the SOCP method, the transmit power values obtained at each iteration are taken as solver CVX's iterative values.
\section{Conclusions}
In this paper, we study the WSRM optimization problem for a multicell OFDMA multiplexing system. We formulate and propose an SPCA-based convex approximation of the optimization problem, which is known to be non-convex and NP-hard. This iterative SOCP optimization is provably convergent to the local optimal solution. Some numerical issues related to the algorithm implementation are also discussed. Particularly, in terms of the convergence rate, this algorithm exhibits excellent performance and outperforms some previously analyzed solutions to the WSRM optimization problem. For power optimization with SINR constraints, SOCP-based beamforming design is straightforward without using uplink-downlink duality. SOCP optimality conditions may be helpful in performance analysis or in enhancing the design criteria without resorting to the virtual uplink problem.

There are many worthy extensions of this paper. Decentralized beamforming optimization would be an interesting future project. Power minimization under per-user rate constraints rather than per-subcarrier SINR constraints can also be considered. Another possible direction is to consider users with multiple antennas. In that case, the convergence speed is expected to be lower for SOCP-based optimization, for which a trade-off investigation can be performed.

\profile{Mirza Golam Kibria}{received a B.E.\ degree in electronics and communication engineering from Visveswaraiah Technological University (VTU), India, in 2005 and an M.Sc degree in wireless communications from Lund Institute of Technology (LTH), Lund University, Sweden, in 2010. Currently, he is working towards his Ph.D degree at the Graduate School of Informatics, Kyoto University, under the Japanese Government Ministry of Education, Culture, Sports, Science, and Technology (MEXT) Scholarship Program. His research interests include downlink precoding, interference management, and power control in cellular systems.}

\profile{Hidekazu Murata}{received B.E., M.E., and Ph.D.\ degrees in electronic engineering from Kyoto University, Kyoto, Japan, in 1991, 1993, and 2000, respectively. In 1993, he joined the Faculty of Engineering, Kyoto University. From 2002 to 2006, he served as an Associate Professor of the Tokyo Institute of Technology. He has been at Kyoto University since October 2006 and is currently an Associate Professor at the Department of Communications and Computer Engineering in the Graduate School of Informatics. His major research interests include signal processing and its hardware implementation, particularly, its application to cooperative wireless networks with cognitive radio capabilities. He received a Young Researcher's Award from the IEICE of Japan in 1997, the Ericsson Young Scientist Award in 2000, and the Young Scientists' Prize of the Commendation for Science and Technology by the Minister of Education, Culture, Sports, Science, and Technology in 2006. He is a member of the IEEE.}

\profile{Susumu Yoshida}{received B.E., M.E.\, and Ph.D.\ degrees all in electrical engineering from Kyoto University, Kyoto, Japan, in 1971, 1973, and 1978, respectively. Since 1973, he has been with the Faculty of Engineering, Kyoto University, and he is currently a full professor of the Graduate School of Informatics, Kyoto University. During the last 30 years, he has been mainly engaged in research on wireless personal communications. His current research interests include highly spectrally efficient wireless transmission techniques and wireless ad hoc networks. From 1990 to 1991, he was a visiting scholar at WINLAB, Rutgers University, U.S.A.,\ and Carleton University in Ottawa. He served as a TPC Chair of IEEE VTC 2000-Spring, Tokyo, and a General Chair of IEEE VTC 2011-Spring, Yokohama. He was a guest editor of IEEE J-SAC on Wireless Local Communications published in April and May 1996. He received the IEICE Achievement Award and Ericsson Telecommunication Award in 1993 and 2007, respectively.}
\end{document}